# A molecular dynamics investigation of the mechanical properties of graphene nanochain

Yongping Zheng[a] Lanqing Xu[a] Zheyong Fan[b] Ning Wei[b] and Zhigao Huang[a]

[a] *School of Physics and OptoElectronics Technology,*
*Fujian Normal University, Fuzhou 350007,*
*China;* [b] *Department of Physics, Xiamen University, Xiamen 361005, China*

## Abstract

In this paper, we investigate, by molecular dynamics simulations, the mechanical properties of a new carbon nanostructure, termed graphene nanochain, constructed by sewing up pristine or twisted graphene nanoribbons (GNRs) and interlocking the obtained nanorings. The obtained tensile strength of defect-free nanochain is a little lower than that of pristine GNRs and the fracture point is earlier than that of the GNRs. The effects of length, width and twist angle of the constituent GNRs on the mechanical performance are analyzed. Furthermore, defect effect is investigated and in some high defect coverage cases, an interesting mechanical strengthening-like behavior is observed. This structure supports the concept of long-cable manufacturing and advanced material design can be achieved by integration of nanochain with other nanocomposites. The technology used to construct the nanochain is experimentally feasible, inspired by the recent demonstrations of atomically precise fabrications of GNRs with complex structures [Phys. Rev. Lett,2009,**102**,205501; Nano Lett., 2010, **10**,4328; Nature,2010,**466**,470]

## I. INTRODUCTION

Carbon-based nanomaterials such as graphene and carbon nanotubes (CNTs) are among the most versatile materials for bottom-up construction of artificial objects in the nanometer-scale.[1–8] Among the various attractive properties the superlative elastic stiffness and ultimate tensile strength possessed by graphene and CNT have sparked intensive research interests.[9–16] Theoretically, density functional theory calculations,[17,18] molecular dynamics (MD) simulations,[19–22] atomistic reaction pathway calculations,[23] and other numerical approaches have been carried out.[24–27] Enormous Young's modulus ($\sim 1.0$ TPa) and tensile strength (a few tens of to one hundred GPa) were observed. Experimentally, the Young's modulus of graphene sheets (less than five layers thick) was reported to be 0.5 TPa[28] and the intrinsic breaking strength of monolayer graphene was identified to be around 123 GPa.[29] The axial Young's modulus of CNTs was measured to be $\sim 1.0$ TPa[30] and the tensile strength of individual multiwall CNTs was determined to be $\sim 60$ GPa.[31] Reinforcing roles of graphene and CNTs played in polymers and polymer matrix have also been well demonstrated.[32,33]

Beyond the examinations of pristine graphene and CNTs, nanodesigns of graphene- and/or CNT-based integrated architecture have attracted significant attentions for the successful implementation of the 'bottom-up' strategy. This concept gives rise to not only grander systems for large scale applications but also fascinating new properties due to the structure engineering involved in. CNTs were joined in X shape or Y shape or circle to form crossbar,[34] super-graphene,[35] super CNT,[36] super cubic or diamond,[37] nanorings,[38] and other types of networks.[39] Meanwhile, graphene sheets and nanoribbons were assembled into hierarchical morphology,[40] forged into network,[41] folded into grafold,[42,43] joined into Möbius strips,[44] patched or defected into stitched graphene,[45–47] etc. Moreover, graphene and CNTs were joined to create pillared-graphene[48] or other types of similar conformations.[49] CNT-based space elevator cable[50] and super-bridges suspended over CNT cables[51] were investigated to explore the application possibilities in mega- or kilometer length-scale. In these reports the super-high mechanical strength of graphene and CNTs was found to be retained at least in part in the newly constructed conformation, and intriguing material performance was observed. For example, graphene-CNT integrated pillared-graphene architecture was reported to display enhanced hydrogen storage capacity,[48] high thermal rectification mutability,[52] and linear mechanical response.[53]

Currently, growing large-size monolayer graphene still remains a challenge. Recently, a chem-ical vapor deposition (CVD) technique has been devised to grow large-area, single-layer stitched graphene patchwork.[54] GNRs with different topologies and widths can be precisely fabricated from molecular monomers.[55] In addition, single chains of carbon have been produced to link graphene sheets[56] as well as carbon nanotubes.[57] Moreover, many nanostructures are demonstrated to be built at single atom level through electron beam welding,[58] thermal reduction,[59] CVD,[60,61] etc. Topologically nontrivial configurations could be realized by mono-atomistic control of synthetic assembly, enabling the exploration of new nanoscale architectures.

Most of the aforementioned architectures are constructed by joining separate parts through carbon-carbon bonds, i.e., it is bonding interaction that links the various parts together. Very recently, the authors proposed a nanostructure constructed through non-bond linking, called knitted graphene, to produce large area 'graphene sheets' and observed mechanical robustness and high tensile strength within this architecture.[62] In this work we invoke the concept of non-bond linking and propose a new architecture, termed graphene nanochain, created from pristine or twisted graphene nanoribbons. We aim to achieve a mechanically stable and strong structure out of flexible GNR strips which could be manufactured in large-scale to produce long cable. Mechanical properties of the nanochain are investigated via molecular dynamics simulations, and parametric studies are carried out to address the length, width, torsion and defect effects. Interesting new properties might be attained by integrating the chain with other nanocomposites further.

## II. MODELS AND METHODS

### A. Models

To be specific, we only consider nanochains constructed from zigzag graphene nanoribbons (ZGNRs), although the same methods can be extended to the case of armchair GNRs. Firstly, a monolayer H-terminated ZGNR with length $L$ and width $W$ is rolled up to create a nanoring (joined by the two armchair extremities, Fig. 1(a,b)). Then, the created nanorings are interlocked one by one to generate the first type of graphene nanochain (TYPE I, Fig. 1(c)). Furthermore, we twist the ribbon by an angle of $N_t \times 180º$ ($N_t$ being integers) before it was rolled up and jointed at the short (armchair) extremities to form a Möbius-like nanoring, and then interlock these Möbius-rings to construct the second type of graphene nanochain (TYPE II, Fig. 1(d)). After structural relaxation the chain transformed from its initial saddle-like conformation into 'helix' configuration, which is believed to be more homogeneous and stable with lower energy. To facilitate later discussions, we divide a constituent nanoring in the nanochain into two regions: the junction-region and the GNR-region (Fig. 1(c)). We will study systematically the size effect by varying the length $L$ and width $W$ of the constituent GNRs and study the strain effect by considering different twisting angles for TYPE II nanochain. Furthermore, defect effects are also investigated in view of the possible presence of defects in real experimental situations.

### B. Computational methods

MD simulations are carried out using LAMMPS[63] MD package with the AIREBO potential[64]. The parameters of the REBO part of the potential are set as suggested[65] to terminate the unphysical high bond force arising from artificial switching functions. The time step is chosen to be 0.1fs. Structural relaxation is performed using conjugated gradient algorithm.[66] During tensile tests NPT ensemble is adopted (300K, temperature control by Nosé-Hoover thermostat[67]) and the systems are deformed along the longitudinal direction of the chain ($z$-axis) at a rate of 0.001/$ps$. Stresses and strains are computed every 1000 MD steps. A damping parameter of 2.0 is introduced to dissipate the undesired oscillations during the simulation. Periodic boundary conditions are implemented along the $z$-axis only, and vacuum spaces of 20 Å are introduced in the other two directions ($x$- and $y$-axis) to ensure that there is no external effect.

The mechanical properties are mainly revealed by the stress-strain relation, where the strain $\varepsilon$ under tension is define as

$$\varepsilon = \frac{L - L_0}{L_0} = \frac{\Delta L}{L_0}, \qquad (1)$$

where $L_0$ and $L$ are respectively the lengths of the entire nanochain before and after stretching. The soft nature of the chain results in diversity in its equilibrium conformation, and we take $L_0$ as the maximum length of the nanochain among various energy minimum states. Other related parameters, such as ultimate tensile strength $\sigma_c$, fracture strain $\varepsilon_F$, and the maximum force $F_c$ will also be discussed. However, Young's modulus is not analyzed, since we only study the mechanical performance under heavy load.

In order to calculate the stress-strain relations during deformation, the per-atom stress tensor for each individual carbon atom is first calculated[63,68–70]

$$\sigma_{ij}^{\alpha} = \frac{1}{\Omega^{\alpha}} \left( \frac{1}{2} m^{\alpha} v_i^{\alpha} v_j^{\alpha} + \frac{1}{2} \sum_{\beta=1,n} f_i^{\alpha\beta} r_j^{\alpha\beta} \right), \tag{2}$$

where $\alpha$ and $\beta$ are atomic indices and $i$ and $j$ are Cartesian indices. $\Omega$ is the representative volume under stress, $m$ the mass, $v$ the velocity, $f$ and $r$ the force and distance between two atoms. The total stress can then be computed by averaging the atomic stresses over all the atoms in the system. Fluctuations are smoothed by averaging the results over the latter 300 MD steps of the relaxation period. Due to thermodynamic fluctuations, the instant cross-sectional area of the nanochain is difficult to measure, and we calculate the volume by assuming a uniform cross-sectional area of $W^2$ (Fig. 2). In section III C the strength is also reported as the force $F = \sigma A$, which is independent of the attributed cross-sectional area.

Our simulation method is validated by calculating $\sigma_c$ and $\varepsilon_F$ of a 300Å×60Å pristine graphene sheet stressed along the zigzag direction. To verify the results with other literature reports, the stress is computed by defining the thickness of graphene as 3.35 Å. We obtained a tensile strength of 106 GPa and a fracture strain of 0.205, which agree well with the experimental measurements as well as other theoretical reports as listed in Table I.

## III. RESULTS AND DISCUSSIONS

### A. Length dependence of mechanical properties

In this section, we investigate the effect of length of the constituent GNRs on the mechanical properties of the resulting nanochains. To this end, five cases of TYPE I graphene nanochains are constructed; each has four identical rings interlocked together. We have enlarged the number of rings to more than 60 and found negligible differences in the derived results. The length $L$ of the graphene nanoribbons used to create the corresponding nanorings ranges from 160 Å to 320 Å.

The main results are presented in Fig. 3(a). The cross-sectional areas of both nanochains and GNRs are defined as $W^2$ (Fig. 2(a,b)) and this rule is always adopted hereafter. The first noticeable observation is that the ultimate strengths of nanochains with different GNR lengths are comparable to each other, which are about 17.5 GPa, a little lower than those of pristine GNRs, which are about 21.2 GPa (This value is about a factor of $3.35/W$ smaller than the one as reported in section II B, since we take the cross-sectional area to be $W^2$ here and $3.35W$ there). The fact that the ultimate tensile strength of the nanochain is independent of the length of the constituent GNRs indicates that the junction-region plays a major role in the fracture process, which will be discussed in detail later.

Another important observation is that $\varepsilon_F$ of the nanochain decreases (nearly linearly) with the increaseing of GNR length, in sharp contrast with the case of GNR, where $\varepsilon_F$ does not depend on the length (Fig. 3(b,c)). This difference can be understood as follows. For pristine GNR, the whole system is homogeneous. Since the elastic constants are the intrinsic properties of a material, they shouldn't vary with size when the analyzed sample is large enough.[21] For the nanochains, as pointed out earlier, it is convenient to divide one constituent nanoring into a junction-region and a GNR-region, with original lengths $L_{0junc}$ and $L_{0GNR}$ respectively. the junction region has nearly the same length for nanorings with different lengths, while the GNR-region of a longer nanoring is longer than that of a short ring. In the whole deformation process up to failure, the individual elongation length of the junction-region $\Delta L_{junc}$ are nearly the same for Nanochain-L160 and Nanochain-L320, and that for the GNR-region $\Delta L_{GNR}$ is approximately proportional to the original length of the GNR-region, as demonstrated in Fig.4. This indicates that both of these two regions have a definite individual strain, which can be denoted as $\varepsilon_{junc}$ and $\varepsilon_{GNR}$, respectively. Thus, the effective strain of the nanochain can be expressed as

$$\varepsilon = \frac{L_{0junc}\varepsilon_{junc} + L_{0GNR}\varepsilon_{GNR}}{L_{0junc} + L_{0GNR}} = \varepsilon_{junc}\frac{1 + \frac{\varepsilon_{GNR}}{\varepsilon_{junc}}\frac{L_{0GNR}}{L_{0junc}}}{1 + \frac{L_{0GNR}}{L_{0junc}}}. \tag{3}$$

Since the junction-region elongates faster than the GNR-region does, i.e., $\varepsilon_{junc} > \varepsilon_{GNR}$, it follows from the above equation that $\varepsilon$ decreases with the increasing of $L_{0GNR}$, as claimed.

Lastly, it can be noted that the stress-strain relations for pristine GNRs are nonlinear, which can be attributed to the anharmonic terms in the interaction potential. In contrast, the stress-strain relations for nanochains consist of two segments, a shorter one with lower slope in the small strain region and a longer one with higher slope in the large strain region. In the small strain region, the atoms around the soft links would rearrange their positions and dissipate part of external load, resulting in a slower increasing of stress. The nearly linear stress-strain response at the higher strain region can be ascribed to the appearance of the junctions, where the in-plane $sp^2$-bonds are transformed into off-plane structures with deteriorated $\pi$ bonds. Similar linear stress-strain relation has also been observed in pillared-graphene architecture where $sp^2$-bonds also exhibit an off-plane structure.[53]

## B. Spatial stress distributions

We now turn to analyze the spatial distribution of the atomic stresses, which is important since it reveals the failure mechanisms underpinning plastic deformations. Figure 5 depicts a vivid scenario of the temporal evolution of atom positions and per-atom stresses during tension. Firstly, stresses are accumulated at the four corners of the junctions (as shown by red atoms in Fig. 5(a)) and the stresses in the GNR-regions are relatively uniform and much lower. Then, the first bond breakage initiates at one of the corners, as illustrated in Fig. 5(b). Finally, a crack propagation is nucleated through bonds bearing the maximum loads. Plastic deformation is found to proceed via tearing of the belt, as can be seen in Fig. 5(c) and (d). These observations suggests that the mechanical performance of the nanochain is mainly determined by the ligature atoms at the corners of the junctions (see also Fig. 6).

Junctions are the place where two links meet and power comes to the links mainly through the edges. Thus the maximum force will appear among the ligature atoms, especially the four corners. Those atoms are the effective ligature atoms responsible for the fracture. Furthermore, the strongly coupled stretching and bending of the chemical bonds can weak the system and impel the junctions to break earlier. The junctions obviously deteriorated the maximum force and the ultimate strength the chain can sustain under tension, as well as the ductility, as have been seen in Fig. 3 (b) and (c).

## C. Width dependence of mechanical properties

We now proceed with the size effect. An increase in ribbon width serves to increase the effective cross-sectional area as well as the available surface area which is desirable for multiple composite design such as adhesion. The ultimate strengths of nanochains with widths of 15 Å and 25 Å can be found to be approximately 18.1 GPa and 7.2 GPa, respectively (Fig. 7(a)). At first sight one may arrives at a counter-intuitive conclusion that narrow chain possess superior mechanical strength. But it is noteworthy to observe that the strength of GNR is also degraded by 7.0 GPa with the ribbon width increased from 15 Å to 25 Å. The significant loss in the maximum stress comes mainly from the larger cross-sectional area of the wider nanochain. To circumvent this ambiguity, we discuss the width effect in terms of force (Fig. 7(b)), for which the definition of the cross-sectional area is unnecessary. The obtained maximum forces are 43.0 nN and 71.5 nN for GNRs with widths 15 Å and 25 Å, respectively, which are proportional to their corresponding ribbon widths. In contrast, the simulation gives negligible difference of maximum forces between wide and norrow nanochains. An increase of ribbon width from 15 Å to 25 Å yields about 1.8% depression in the maximum force (38.7 nN for Nanochain-W15 and 38.0 nN for Nanochain-W25), which indicates nearly equivalent load carrying capacities for both nanochains.

To gain further insight into the width effect, we calculate $\varepsilon_F$, $\sigma_c$ and $F_c$ for nanochains and GNRs with ribbon widths varying from 15 Å to 50 Å, as shown in Fig. 7(c-e). The fracture strains $\varepsilon_F$ for GNRs fluctuate between 0.186 and 0.204, and those for nanochains fluctuate between 0.115 and 0.153 (Fig. 7(c)). The ultimate strength $\sigma_c$ for both nanochains and GNRs decrease with the increasing of the ribbon width, as expected.(Fig. 7(d)). As for the maximum force (Fig. 7(e)), it increases linearly with the increasing of ribbon width for GNRs, while for nanochains, a peak value of 55.1 nN is found with ribbon width of 35 Å, in which case it is found that there are more effective ligature atoms sharing the loads. This result suggests that nanochains with medium thickness may possess the highest load carrying capacity.

### D. Strain effect

It has been proposed that twisted ribbons could endure significantly larger tensile strain due to intrinsic unwinding of the twisted structure subjected to load[71]. In this section, the effect of strain resulted from twisting the GNRs of nanochain is investigated by studying the mechanical properties of untwisted TYPE I nanochain ($N_t = 0$) and twisted TYPE II nanochains with different twisting angles ($N_t = 2$, 4 and 6). We have chosen GNRs with relatively narrow width (15 Å) as components of the nanochains, since it has been previously found that as the width increases, it becomes more and more difficult to synthesize the Möbius structures by GNRs.[72] Under tension the links transformed from twisted to helical conformations to dissipate external loads, which has also been observed in other systems.[70,73] The deformation leads to anomalous twisted and folded conformation resulting from the increased curvature (Fig. 8). An important question then arises as to whether arbitrary rotation causes a reinforcement, or a destruction, or only negligible effect to the composite.

Fig. 9(a) shows the stress-strain relations for nanochains with $N_t$ = 0, 2, 4 and 6, from which we can see that the larger the twisting angle, the slower the increase of stress at the beginning of the stretching, and the larger the fracture strain. For twisted nanochains, external load can be dissipated through atom position rearrangement in both junction-regions and GNR-regions. Moreover, we also observe a significant reduction of the equilibrium length for twisted nanochains. Therefore, nanochain with larger twisting angle has more space for atomic rearrangement, and exhibits a lower rate of stress increment in the light load region and develops a larger fracture stain. Note that if one takes the equilibrium lengths of the nanochains uniformly as that of the untwisted nanochain, one would find that the fracture strains of nanochains with larger twisting angles are lower than those of nanochains with smaller twisting angles (Fig. 10). The non-zero stresses at $\varepsilon = 0$ for the $N_t > 0$ cases show that these chains are pre-strained. Thus when stretched to the same level the twisted nanochains are much easier to break down. We have performed simulations at other engineering strain rates and reached the same conclusion.

While twisted nanochains are more ductile, their ultimate strengths are reduced by a few percentages compared to that of the untwisted nanochain, as can be seen from Fig. 9(b). The ultimate strengths of nanochains decrease from 18.21 GPa to 16.82 GPa with $N_t$ varying from 0 to 6. For twisted nanochains, ribbons in the junction-regions not only bend but also coiled up, which generates pre-strains and deteriorates the local bonds of the ligature atoms. In section III B, we have demonstrated that the strength of the system depends largely on the strength of the ligature atoms. This explains the reduction of the ultimate strength for the twisted nanochains.

### E. Defect dependence of mechanical properties

In view of the presence of the unavoidable defects during the experimental fabrication process, we consider the effects of defects, both regular and random, on the mechanical performance of the nanochains and GNRs. In the regular defect case, single defect aligning parallel/perpendicular to the elongation direction is considered and the defect size is measured by the number of missed atoms (Fig. 11(a,b)). In the random defect case, single-atom defects are randomly distributed in the investigated regions.

Firstly, we study the effects of regular defects. From the results as presented in Fig. 11(c), one can see that the parallel defect acts like a random disturbance to the stabilities of both the nanochain and the GNR structures. The fracture strains and ultimate stresses of GNRs are reduced by about 30% and 20% respectively, with small fluctuations against the defect size. The depressions of the ultimate strains and stresses for nanochains with parallel defects are less than 20%, with larger fluctuations. For the case of perpendicular defect, both the fracture strain and the ultimate strength of GNR decrease as the defect size increases. In contrast, the nanochain is relatively insensitive to this kind of defect and only a random drop less than 10% is found for both fracture strain and ultimate strength. These phenomena could be understood from the following point of view. In GNR the parallel defect destroys the bonds parallel to the deformation, which only causes a random intrusion to weak the system. However, the perpendicular defect destroys the bonds perpendicular to the deformation which sustain the majority of the load, leading to more depreciated mechanical strength. For nanochain, mechanical performance depends largely on how strong those ligature atoms are. Since none of these defects exist at the junctions, their appearance can only lead to random conformation perturbations, resulting in a small overall loss in the robustness of the system. Temporal evolutions of the cracking procedures can further complement the above explanations. Pristine GNRs are found to rift from regions containing defects, whereas the nanochains are observed to rupture from the junctions where the ligature atoms reside in. From the above results one can come to a conclusion that regular defects have minor influence on the mechanical properties of nanochains.

Secondly, we investigate the effects of random defects since the occurrence of defects is hard to control in the manufacturing process. Values of $\varepsilon_F/\varepsilon_{F0}$ and $\sigma/\sigma_0$ for nanochains and GNRs obtained from extensive MD simulations for different random defect numbers $N_D$ are displayed in Fig. 12. The error bars are derived from ten simulations of statistically independent realizations of systems with given defect numbers. From Fig. 12, one can see that GNRs present a downgrade trend in both $\varepsilon_F/\varepsilon_{F0}$ and $\sigma/\sigma_0$. On the contrary, $\varepsilon_F/\varepsilon_{F0}$ of the nanochains present a degrading-improving trend as $N_D$ increases, and $\sigma/\sigma_0$ degenerates at first and increases a little when $N_D >$ 300. The earlier reductions in both $\varepsilon_F/\varepsilon_{F0}$ and $\sigma/\sigma_0$ should be ascribed to the defect-induced weakening of the architecture. The failure is also more abrupt (brittle-like) when $N_D < 300$. However, for nanochains with $N_D > 300$, we found that the chain does not rupture quickly after the first breakage of the bond and a strengthening-like behavior is observed (Fig. 13). Some of the chains are found to have repeated 'increase-depress' trend of stress-strain evolutions. The ultimate stress is improved by a little amount, and the fracture point is much later than the low defect coverage cases, which indicates that the nanochain become more ductile and less brittle with high defect coverage. With a close examination we observed that at the junctions $sp^3$-hybridizations are formed by atoms from two adjacent links (Fig.14). It has been previously pointed out that $sp^3$ interwall bonding in defective multiwall carbon nanotubes have strengths exceeding those of single-wall carbon nanotubes containing the same size of initial intrawall defects.[74] In our case the interlink $sp^3$ bonds also play a strengthening role. Nanochains with mechanical strengthening-like behavior are all observed to have interlink $sp^3$ bonds, and with more $sp^3$ bonds the chain is strengthened more.

Finally, we investigate the effect of defect site. From the listed values in Table II, one can see that for one single-atom defect in the junction-region, the fracture strain drops to $\varepsilon_F \approx 0.159$ (Nanochain II), whereas for one single-atom defect in the GNR-region, the fracture strain is 0.197 (Nanochain III), which is nearly equal to $\varepsilon_{F0}$. Low defect level within the junction-region deteriorates the nanochain whereas sparse defects in the GNR-region has only limited influence on the system. With high defect coverage randomly distributed in the entire chain the fracture strain increases to $\varepsilon_F \approx 0.268$ with associated $\sigma_c \approx 2.813$ GPa (Nanochain IV). To understand the origin of the high $\varepsilon_F$ and ductility we remove away the defects in the GNR-region to create Nanochain V and find that $\varepsilon_F$ and $\sigma_c$ of this chain still resemble those of Nanochain IV. Moreover, by removing away defects in the junction-region we construct Nanochain VI, which is found to possess reduced $\varepsilon_F$ of 0.189 and $\sigma_c$ of 2.212 GPa. High defect coverage in the junction-region results in observable improvement in ductility whereas high defect coverage in GNR-region leads to significant depression in the mechanical performance.

## IV. CONCLUSIONS

By molecular dynamics simulations, we demonstrated that joining both extremes of graphene nanoribbons together and interlocking the obtained rings can lead to stable novel architecture, termed graphene nanochain, with ultimate strength a little lower than that of pristine graphene nanoribbons. The parametric studies show that the length of the ribbons have negligible influence on the mechanical strength, whereas the width of the ribbons have significant impacts. Twisting the ribbon induces additional strain which would result in deteriorated strength but improved ductility. The strength of the system is determined mainly by the strength of the ligature atoms. Random and regular defects in the system could affect more or less the failure process. Under heavy random defects situation an interesting mechanical strengthening-like behavior is observed in some nanochains and the ductility of the material is better than GNR. Further experimental investigations are therefore called for to verify these phenomena. The uncovered unique physical properties are found to be independent of the total chain length ranging from several tens to several hundreds of nanometers, which implies that the strength of the nanochain may be re-tained in even longer chains, making a mechanically robust and long cable plausible. Moreover, this conformation possesses omnidirectional flexibility and could be further integrated with other nanocomposites to further expands its diversity and adaptability. The knowledge of the mechanical performance of graphene nanochains as gained here render nanochain a promising candidate for carbon-based functional material in large-scale mechanical applications.

**FIGURES**

TABLES

REFERENCES

---

**FIGURES**

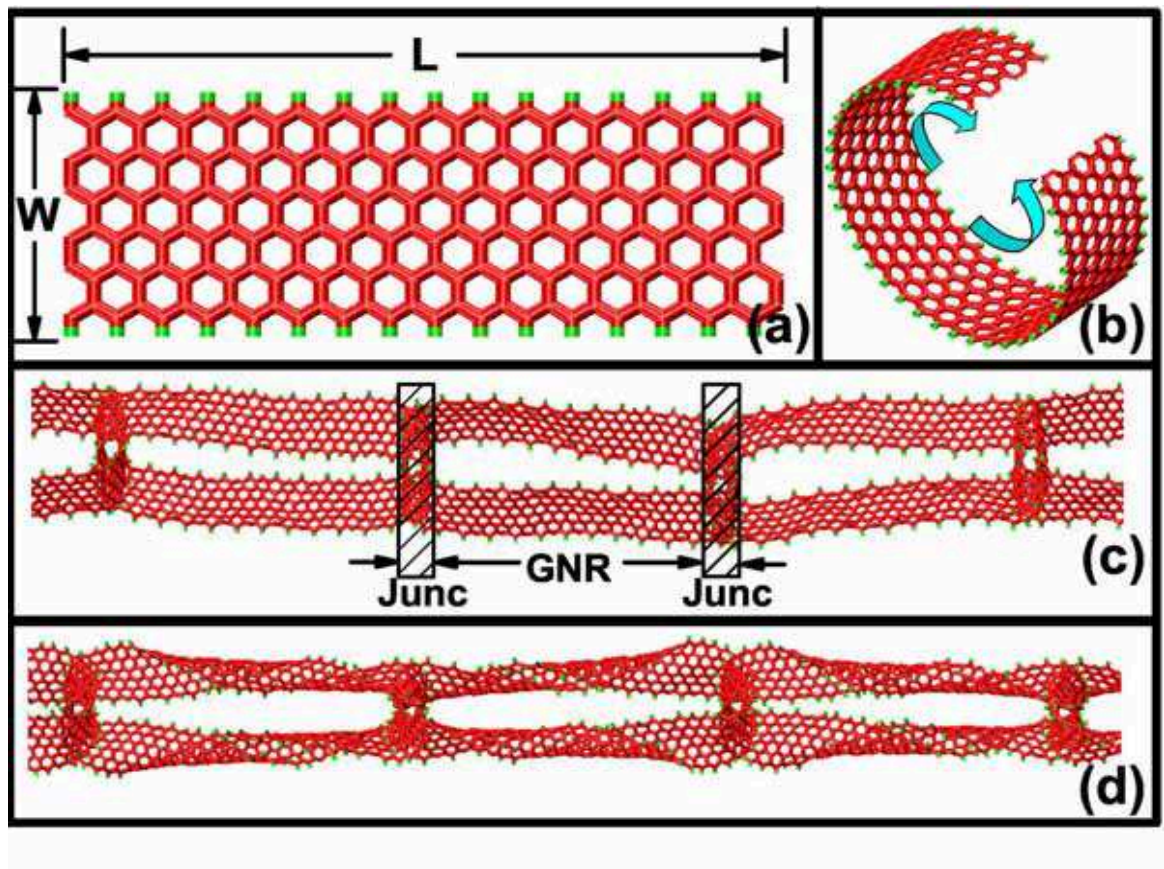

**FIG. 1** (color on line) (a) Zigzag GNR with length $L$ and width $W$, passivated by hydrogen atoms. (b) Rolling up the GNR and joining the two armchair ends to create a nanoring. (c) Interlocking the created nanorings to construct TYPE I nanochain. Junction-regions and GNR-regions are also illustrated. (d) Snapshot of TYPE II nanochain.

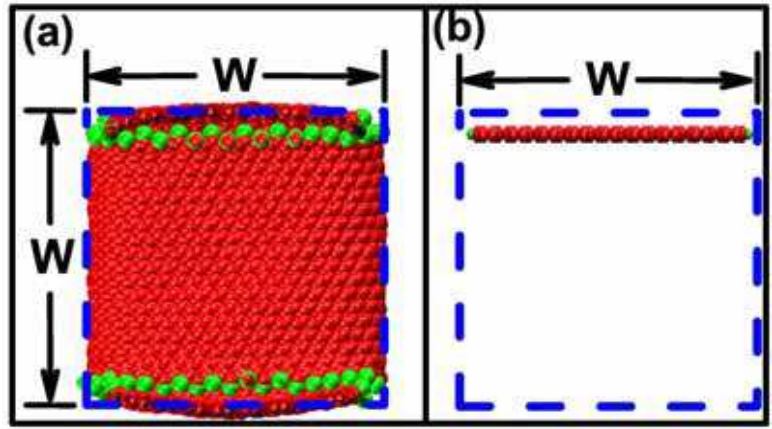

**FIG. 2** (color on line) Definitions of the cross-sectional area for stress computations in section III. (a) nanochain, (b) GNR. $W$ equals to the width of the ribbon used to construct the nanochain and $W^2$ is an approximation corresponding to the cross-sectional area of the nanochain since the real cross-sectional area fluctuates with time.

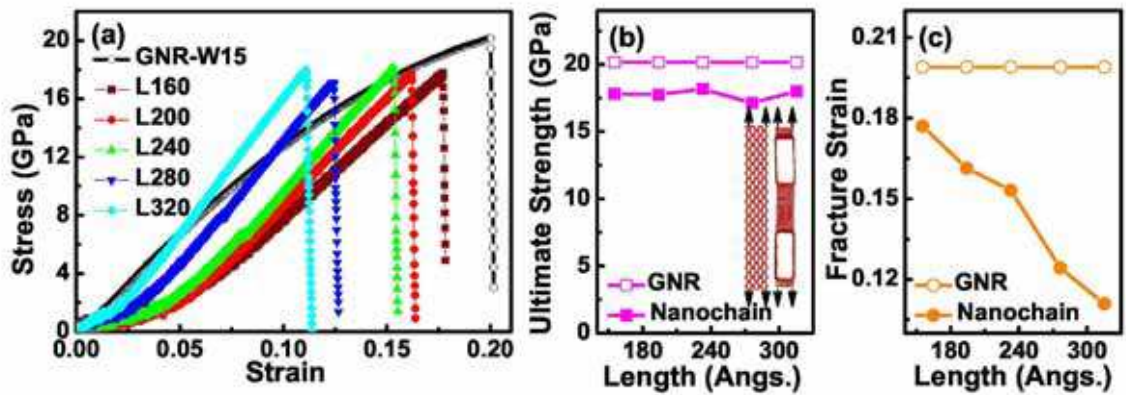

**FIG. 3** (color on line) Length-dependent mechanical properties of graphene nanochains and GNRs. (a) Stress-strain evolutions of TYPE I graphene nanochains with different lengths, where the legend $L\#\#$ denotes the length of GNR used to create one single link. Results from five pristine GNRs with lengths varying from 160 Å to 320 Å are also presented for comparison. The width $W$ of the nanoribbons used to create the links and the width $W$ of the pristine GNRs are both 15 Å. The cross-sectional area in both cases are $W^2$. (b) Ultimate tensile strength with respect to ribbon length. (c) Fracture strain with respect to ribbon length. The insets in (b) illustrate the stretching outlooks.

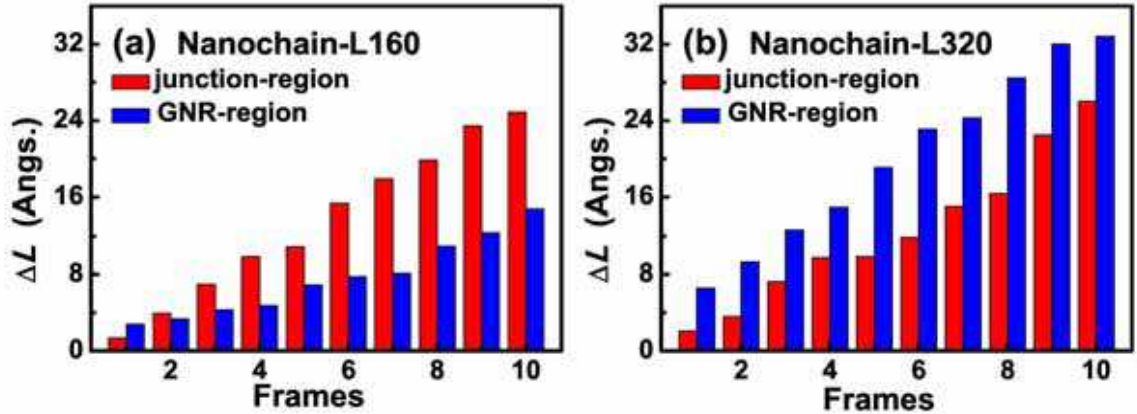

**FIG. 4** (color on line) Absolute elongated length ΔL of the junction-region and GNR-region for (a) Nanochain-L160 and (b) Nanochain-L320. Frames 1 to 10 are evenly selected from the beginning to the end of the deformation procedure.

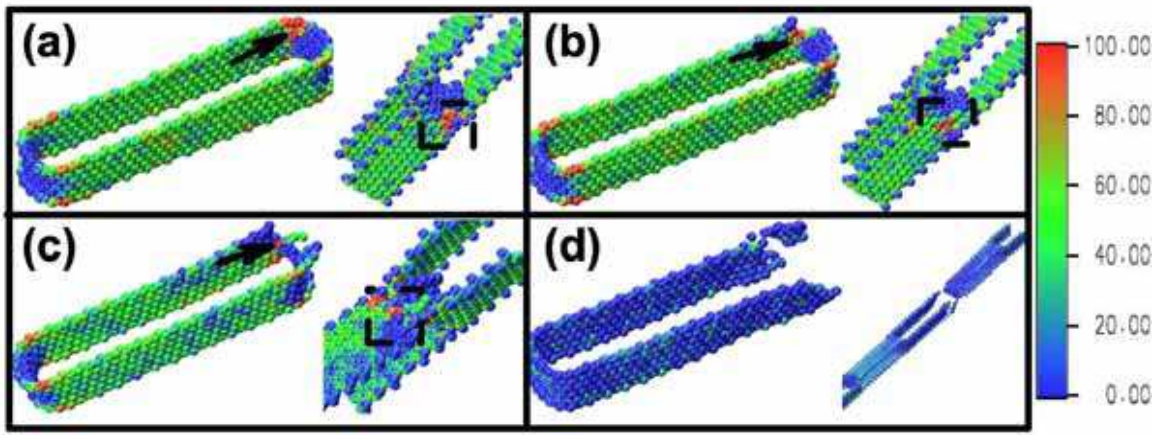

**FIG. 5** (color on line) Evolution of spatial atomic stress distributions in a typical graphene nanochain architecture under tension. In each subfigure the left panel illustrates stress distribution on a single ring (hydrogen atoms at the edges are omitted for clarity) and the right panel demonstrates the stress distribution around the pertinent junction. The positions highlighted by black rectangles are the same as those directed by the black arrows. (a) Before bond breaking, (b) nucleation of the rupture, (c) crack propagation and (d) fracture of the chain. The strain is applied along the longitudinal direction of the link ($z$-axis). As indicated by the arrows, the first breakage of C-C bond occurs at the junction where two links meet and power comes to.

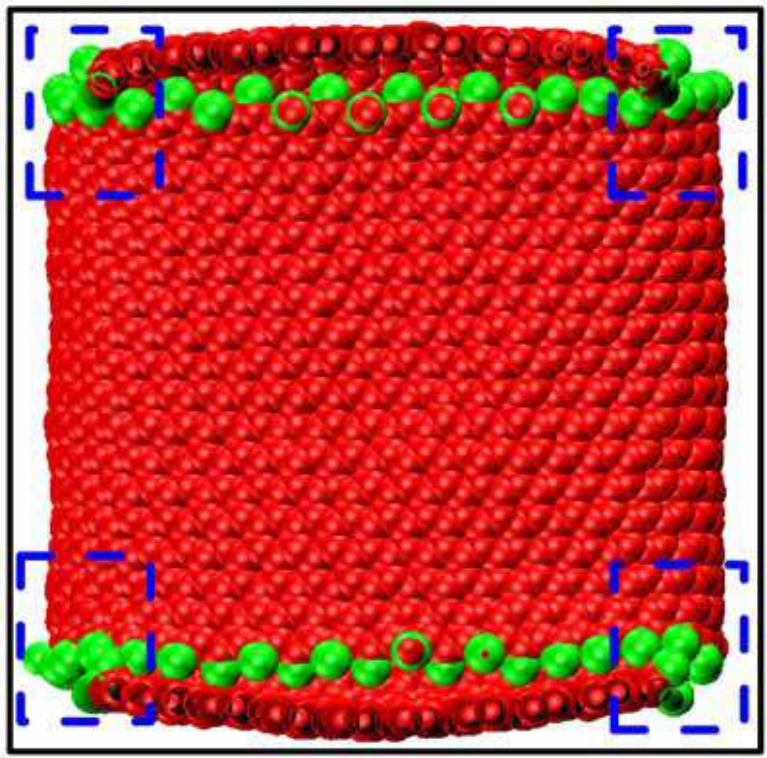

**FIG. 6** (color on line) Illustration of the effective ligature atoms located at the four corners of the junction.

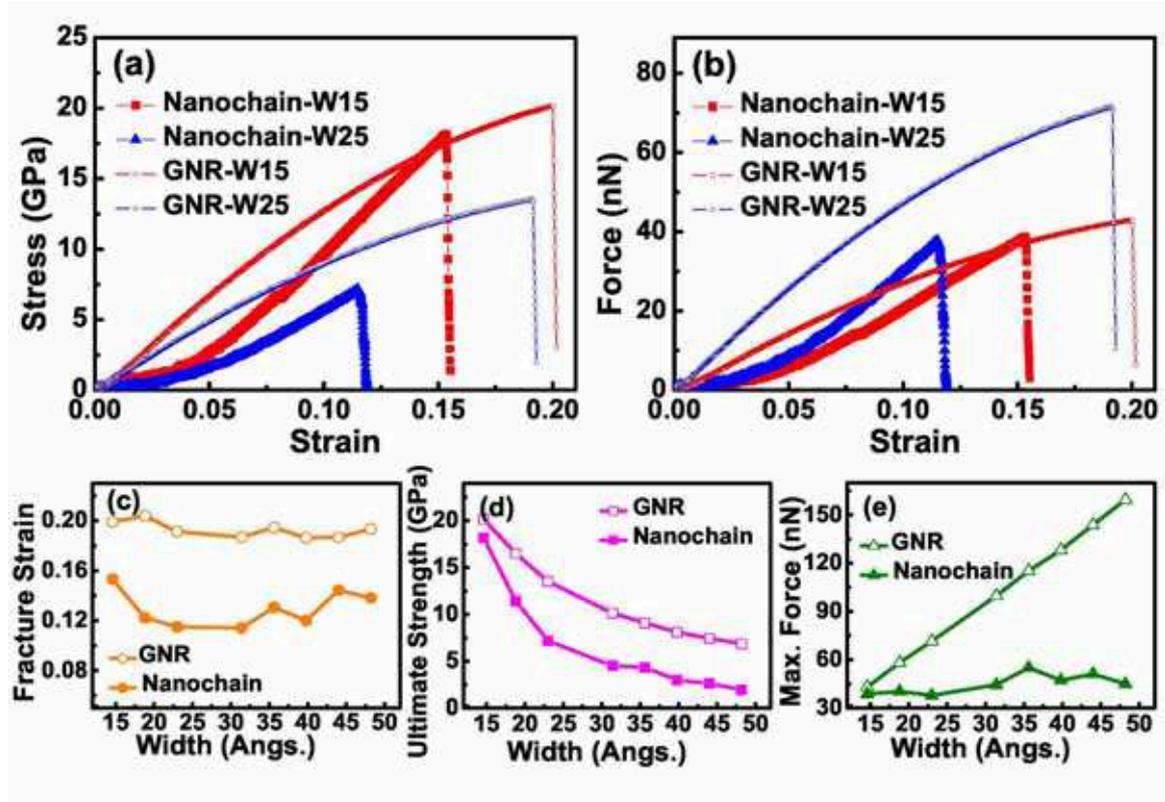

**FIG. 7** (color on line) Width-dependent mechanical properties of graphene nanochains and GNRs. (a) Stress-strain evolutions of graphene nanochains with different widths, where the legend *W##* denotes the width of the GNR used to create one single link. Stress-strain relations of pristine GNRs with the same widths are also presented for comparison. The cross-sectional areas for both circumstances are $W^2$. (b) Force with respect to strain. (c) Fracture strain versus ribbon width. (d) Ultimate tensile strength versus ribbon width. (e) Maximum force at the critical point, as a function of ribbon width.

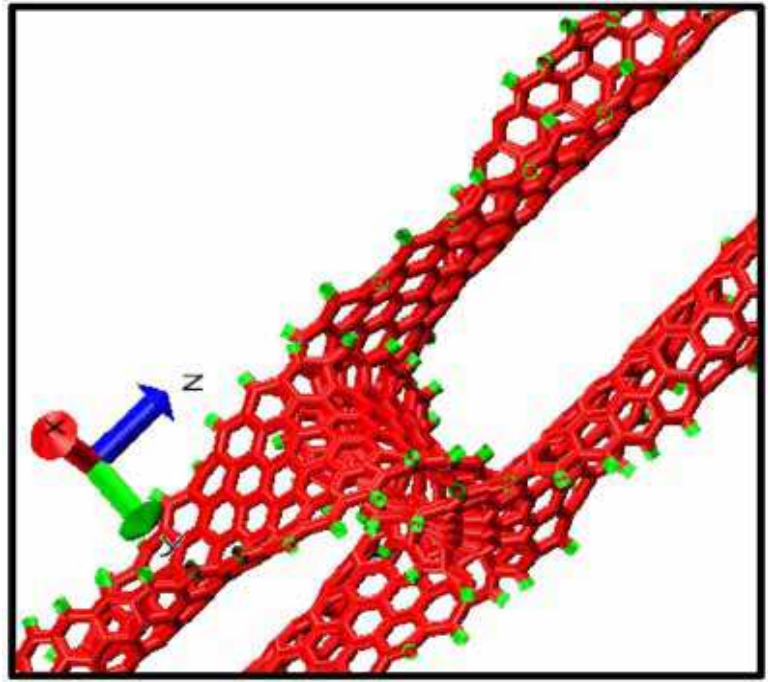

**FIG. 8** (color on line) Snapshot of the bent and folded conformation on the junction. The *z*-axis is the longitudinal direction of the nanochain and loads are acted along this direction.

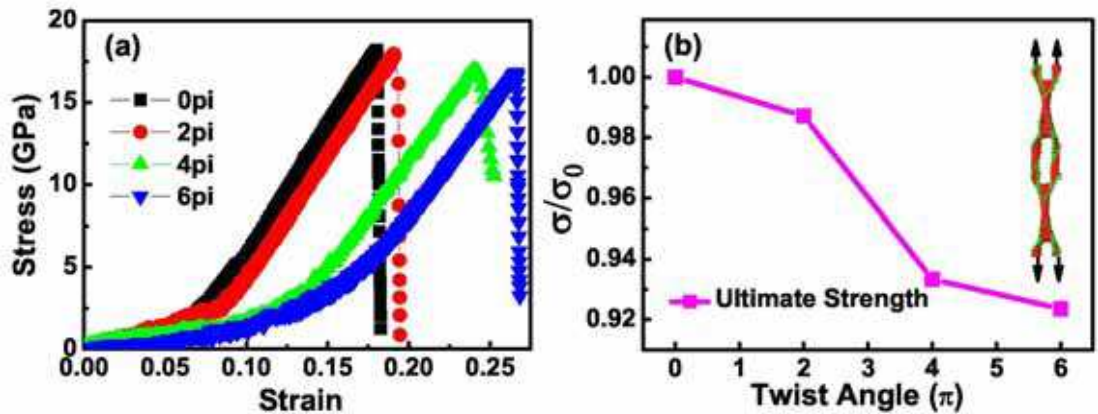

**FIG. 9** (color on line) Effect of strain on the mechanical performance of graphene nanochains. (a) Stress versus strain for TYPE I graphene nanochain (denoted as 0pi) and TYPE II graphene nanochains with $N_t = 2, 4$ and $6$ (denoted as 2pi, 4pi, and 6pi). (b) Relative tensile strength $\sigma/\sigma_0$ with respect to twist angles, where $\sigma_0$ is the ultimate strength of TYPE I graphene nanochain. All four chains are constructed from ribbons with $L = 240$ Å and $W = 15$ Å. Stresses are computed by taking the cross-sectional area as $W^2$. The inset in (b) illustrates the elongation direction.

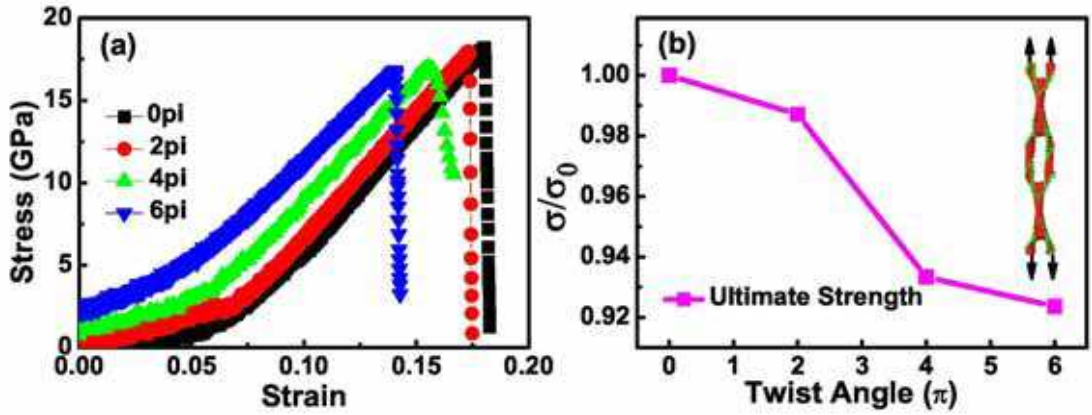

**FIG. 10** (color on line) (a) Stress-strain relations of nanochains with $N_t$ = 0, 2, 4, and 6, where the strains are computed by taking $L_0$ for all the nanochains as that of the untwisted nanochain with $N_t$ = 0. (b) Relative ultimate tensile strength as a function of the twist angle.

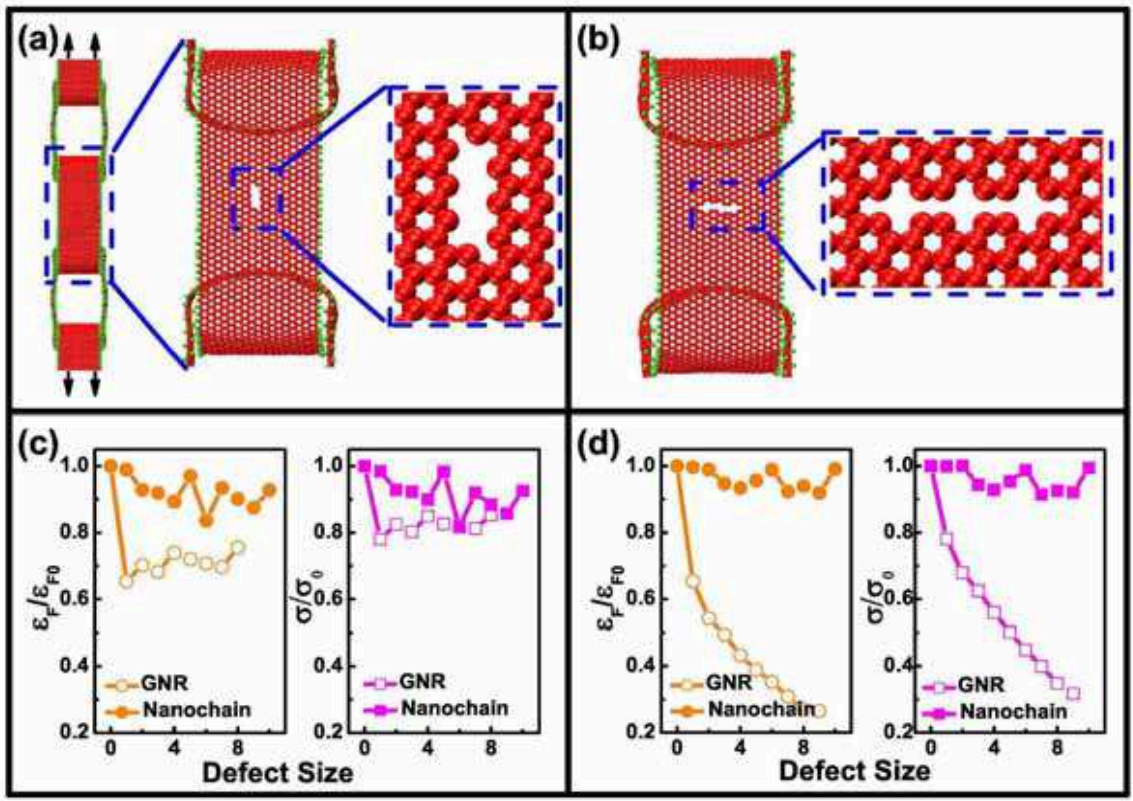

**FIG. 11** (color on line) Effect of regular defects on the mechanical performance of graphene nanochains and GNRs. Tensile tests are loaded along the longitudinal direction. Illustrations of (a) parallel and (b) perpendicular defects in the GNR-region of the nanochain. Variations of relative fracture strain $\varepsilon_F/\varepsilon_{F0}$ and relative ultimate strength $\sigma/\sigma_0$ of nanochains and GNRs with (c) parallel and (d) perpendicular defects, where $\varepsilon_{F0}$ for defect-free nanochain is 0.198, with related $\sigma_0 = 3.855$ GPa. The defect size is measured by the number of missing atoms in the defect. The nanochain subjected to tests are constructed from ribbons with $L = 270$ Å and $W = 40$ Å. The length and width of GNRs are the same as those of nanochains. The cross-sectional area used for stress computation is $W^2$.

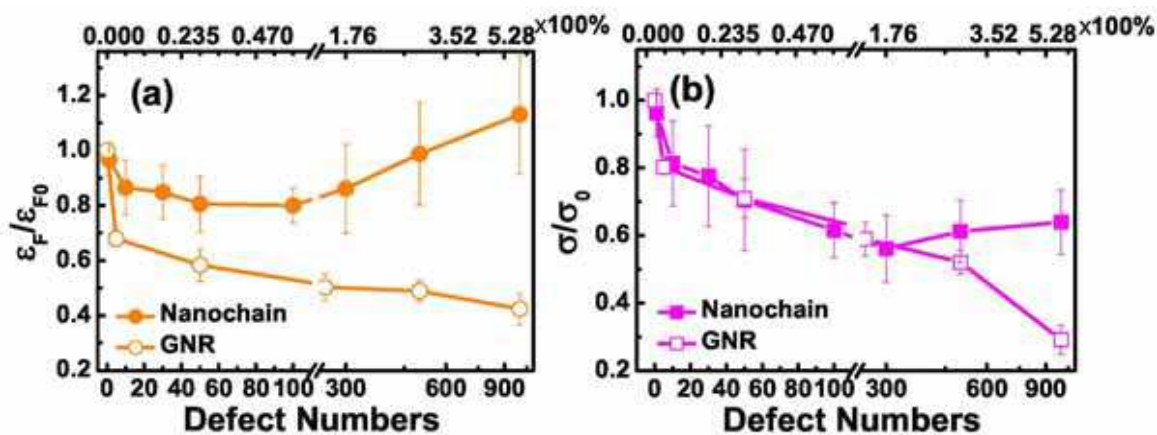

**FIG. 12** (color on line) Variations of (a) relative fracture strain and (b) relative strength normalized by $\varepsilon_{F0}$ and $\sigma_0$, respectively for nanochains and GNRS with respect to the number of random defects $N_D$. Parameters for GNRs and nanochains are the same as those in figure 11. The tick label in the upper axis corresponds to defect coverage (in percentage).

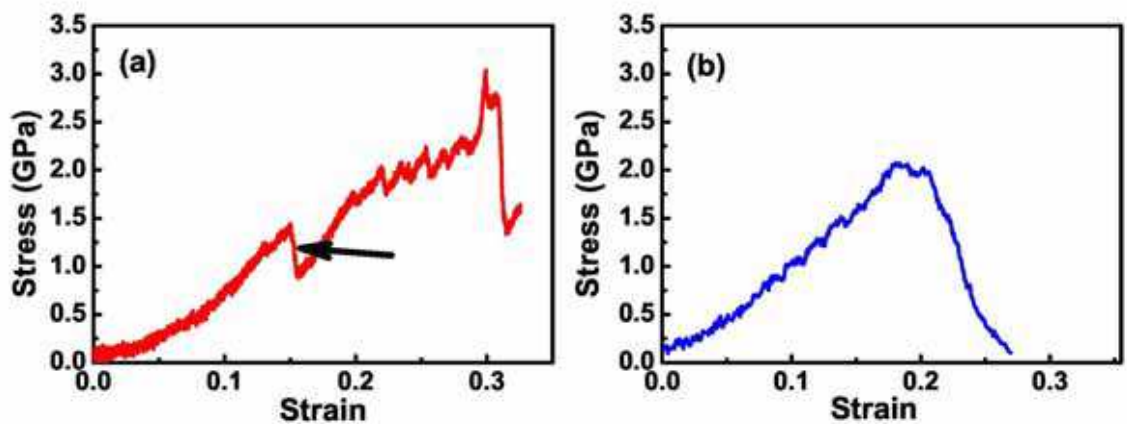

**FIG. 13** (color on line) Stress-strain relations of two nanochains with 1000 random defects. (a) Evolution with a strengthening-like behavior, as indicated by the arrow. (b) Evolution without obvious yield point.

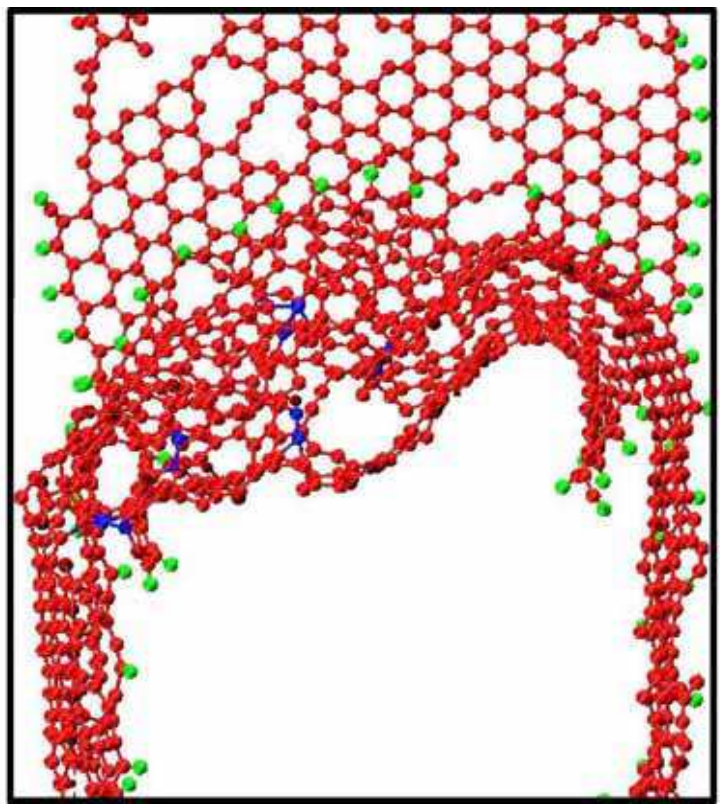

**FIG. 14** (color on line) Snapshot of heavily defected nanochain. Only part of a junction is clipped out for clarify. The blue atoms highlight the interlink $sp^3$ bonds formed by the rearrangement of bonds during the tensile deformation.

**TABLES**

**TABLE I** Mechanical properties of pristine graphene membrane. The results from the present simulation are consistent with previous reports well.

|  | $\sigma_c$ (GPa) | $\varepsilon_F$ | Remarks |
| --- | --- | --- | --- |
| Our Work | 106 | 0.205 | graphene (zigzag direction) |
| Ref.[29] | 130 ± 10 |  | graphene (nano-indenting) |

| | | | $\sigma_c$ (GPa) | $\varepsilon_F$ | |
|---|---|---|---|---|---|
| Ref.[21] | 107 | | | 0.20 | graphene (MD simulations, zigzag direction) |
| Ref.[21] | 90 | | | 0.13 | graphene (MD simulations, armchair direction) |
| Ref.[31] | 60 | | | | SWNT (experiments) |

TABLE II Comparison of fracture strain and ultimate stress between nanochains with different random defect distributions. The structural parameters are the same as those in figure 11.

| | $N_D$ | Defect site | $\sigma_c$ (GPa) | $\varepsilon_F$ | Remarks |
|---|---|---|---|---|---|
| Nanochain I | 0 | | 3.855 | 0.198 | Defect-free |
| Nanochain II | 1 | junction-region | 2.979 | 0.159 | One single-atom defect |
| Nanochain III | 1 | GNR-region | 3.855 | 0.197 | One single-atom defect |
| Nanochain IV | 1000 | Nanochain | 2.813 | 0.268 | Random defects on the entire chain |
| Nanochain V | 356 | junction-region | 2.633 | 0.246 | Same as Nanochain IV, but without the defects in the GNR-region |
| Nanochain VI | 644 | GNR-region | 2.212 | 0.189 | Same as Nanochain IV, but without the defects in the junction-region |